# On the radiative and thermodynamic properties of the cosmic radiations using *COBE* FIRAS instrument data: III. Galactic far-infrared radiation

**Anatoliy I Fisenko, Vladimir Lemberg**

*ONCFEC Inc., 250 Lake Street, Suite 909, St. Catharines, Ontario L2R 5Z4, Canada*

*E-mail: afisenko@oncfec.com*

**Abstract** Using the three-component spectral model describing the FIRAS average continuum spectra, the analytical expressions for the temperature dependences of the thermodynamic and radiative functions of the galactic far-infrared radiation are obtained. The *COBE* FIRAS instrument data in the 0.15 – 2.88 THz frequency interval at the mean temperatures $T$ = 17.72 K, $T$ = 14 K, and $T$ = 6.73 K are used for calculating the radiative and thermodynamic functions, such as the total radiation power per unit area, total energy density, total emissivity, number density of photons, Helmholtz free energy density, entropy density, heat capacity at constant volume and pressure for the warm, intermediate-temperature and very cold components of the Galactic continuum spectra. The generalized Stefan-Boltzmann laws for the warm, intermediate-temperature and very cold components are constructed. The temperature dependence of each component is determined by the formula $I^{\text{S-B}}(T) = \sigma' T^6$. This result is important when we construct the cosmological models of radiative transfer in the inner Galaxy. Within the framework of the three- component spectral model, the total number of photons in our Galaxy and the total radiation power (total luminosity) emitted from the surface of the Galaxy are calculated. Their values are $N_{\text{G total}} = 1.3780 \times 10^{68}$ and $I_{\text{G total}}(T) = 1.0482 \times 10^{36}$ W. Other radiative and thermodynamic properties of the galactic far-infrared radiation (photon gas) for the Galaxy are calculated. The expressions for astrophysical parameters, such as the entropy density/Boltzmann constant, and number density of the galactic far-infrared photons are obtained. We assume that the obtained analytical expressions for thermodynamic and radiative functions may be useful for describing the continuum spectra of the far-infrared radiation for outer galaxies.

## 1. Introduction

It is well-know that Galaxy spectra, such as a continuum, absorption lines, and emission lines provide us with information about the galaxy's velocity and mass, the average age of stellar population, and the star-formation rate (Schneider 2006).

However, it is essential to note that the continuum spectra of the thermal radiation also contain information about the radiative and thermal properties of objects that emitted it. These properties include the total energy density, total radiation power per unit area, total emissivity, number density of photons, free energy density, entropy density, pressure, heat capacity at constant volume etc. It is therefore of interest to study these properties for different types of sky radiations, such as the cosmic microwave background radiation (Mather et al. 1990; Fixen et. al. 1994), the extragalactic far-infrared background (CMB) radiation (Fixen et al. 1998), the galactic far-infrared radiation (Wright et al. 1991; Reach et al. 1995; Fixsen et. al. 1996) and others. As a result, the radiative and thermodynamic properties of the sky thermal radiation can be obtained.

This article is one of a set of articles associated with the study of the radiative and thermodynamic properties of the cosmic radiations using the *COBE* FIRAS observation data. In previous articles, these properties have been discussed in detail for the cosmic microwave background radiation (Fisenko & Lemberg 2014 a) and the extragalactic far-infrared background radiation (Fisenko & Lemberg 2014 b). In the first article were obtained the analytical expressions for the temperature and redshift dependences of the radiative and thermodynamic functions. New astrophysical parameters for the dipole spectrum, such as the entropy density/Boltzmann constant, and the number density of CMB photons were constructed. It was noted the need to consider the dipole contribution to the Stefan-Boltzmann law for the construction of cosmological models. In the second article, the same properties for the extragalactic far-infrared radiation were calculated using numerical methods.

In this article, the *COBE* FIRAS observation data are used to obtain analytical expressions for the thermodynamic and radiative properties of the galactic far-infrared radiation.

In (Wright et al. 1991), the one-component model to describe the FIRAS Galaxy continuous spectrum was proposed. Modeling the spectral emissivity in the form $\nu^{1.65}$, the observed data were fitted using the dust mean temperature $T_{\text{dust}}$ = 23.3 K. As a result, a map of the dust emission was obtained.

Another form for the spectral emissivity $v^2$ has been proposed in (Fixen et al. 1996). The one-component model was used to fit the observed data for Galactic continuous spectrum. In this work, the Galactic radiation treated as a contaminant in the cosmic microwave background radiation.

In (Reach et al. 1995), the Galactic continuum spectra were observed in the wavenumber range from 5 – 96 cm$^{-1}$ using *COBE* FIRAS instrument. To describe the continuum spectra of the galactic far-infrared radiation the three-component model was proposed. This model contains of three components of emission: the warm dust emission, intermediate-temperature dust emission, and emission from very cold dust. The continuous spectrum for each component was fitted using a power-law dependence of the spectral emissivity as $\varepsilon(v) \sim v^2$. This model was used two free parameters (temperature T and optical depth) to fit the observed data.

The present paper focuses on the study of the radiative and thermodynamic properties of the galactic far-infrared radiation. Using a three-component model with a spectral emissivity in the form of $\varepsilon(v) = (v/v_0)^2$, the analytical expressions for the temperature dependences of the total energy density, the total radiation power per unit area, total emissivity, number density of photons, free energy density, entropy density, pressure, heat capacity at constant volume for the warm, intermediate-temperature, and cold components are obtained. These radiative and thermodynamic properties were calculated in a finite range frequencies between 5 cm$^{-1}$ to 96 cm$^{-1}$ with the following mean temperatures: a) warm temperature $T$ = 17.72 K, b) intermediate-temperature $T$ =14 K, and c) very cold temperature $T$ = 6.75 K. The generalized Stefan-Boltzmann laws for the warm, intermediate-temperature and very cold components are constructed. New astrophysical properties for our Galaxy, such as the total number of photons, total radiation power (total luminosity) emitted from a surface of the Galaxy, and others are obtained.

## 2. General relationships

According to (Mather et al. 1994; Reach et al. 1995), the observed continuum spectra $I(\tilde{v})$ of the sky radiation is modeled by four separate components

$$I(\tilde{v}) = B(\tilde{v}, T_0) + \frac{\partial B(\tilde{v}, T)}{\partial T}\Big|_{T=T_0} \Delta T + G_k(l,b) g_k(\tilde{v}) + z(\tilde{v}) Z(l,b) \qquad (1)$$

Here, the first and second terms are the monopole and dipole components of the cosmic microwave background radiation. These components represent the isotropic background. The third term is presented one or two spatial distributions $G_k(l,b)$ with the spectra $g_k(v)$ derived from the all-sky data set. The final term is the zodiacal component with the spectrum described by a power law, $z(\tilde{v}) \propto \tilde{v}^3$. $T_0 = 2.72548$ K is the mean temperature of the CMB radiation (Mather et al. 2013; Fixsen 2009), $\Delta T = T - T_0$ is the temperature anisotropy in a given direction in the sky, and

$$B(\tilde{v},T) = 2hc^2 \frac{\tilde{v}^3}{e^{\frac{h\tilde{v}}{k_B T}} - 1}, \tag{2}$$

is the Planck function at temperature *T*. Here *h* is the Planck constant and *c* is the speed of light and $\tilde{v}$ is the wavenumber (cm$^{-1}$).

According to (Reach et al. 1995), FIRAS data for the galactic spectrum were modeled as a sum of three components (warm, intermediate-temperature, and very cold) of the form

$$I_0(l,b,\tilde{v},T) = G(l,b)\tau_0\varepsilon(\tilde{v})B_{\tilde{v}}(T), \tag{3}$$

where $\tau_0$ is the optical depth at $\tilde{v}_0 = 30$ cm$^{-1}$, and $\varepsilon(\tilde{v},T)$ is the spectral emissivity normalized to unity at $\tilde{v}_0$. In this three-component model, the spectral emissivity for each component has the form

$$\varepsilon(v) = \left(\frac{\tilde{v}}{30\,\text{cm}^{-1}}\right)^2. \tag{4}$$

Using Eq. 4 and Eq. 3, we have an expression for the spectral intensity for the sum of three components as follows

$$I_{0\,G}(l,b,\tilde{v},T) = \varepsilon(\tilde{v})\sum_{i=1}^{3} G_i(l,b)\tau_i B_{\tilde{v}}(T_i) \tag{5}$$

Here $\tau_i$ and $T_i$ are the optical depths and temperatures for the warm (*i*=1), intermediate-temperature (*i*=2), and very cold (*i*=3) components.

According to Eq. 5, the total intensity of galactic far-infrared radiation in the finite range of wavenumbers has the following form:

$$I'_{0\,\mathrm{G}}(\tilde{\nu}_1,\tilde{\nu}_2,T)=\sum_{i=1}^{3}G_i(l,b)\tau_i\int_{\nu_1}^{\nu_2}\varepsilon(\tilde{\nu})B_{\tilde{\nu}}(T_i)\mathrm{d}\tilde{\nu}\,. \qquad (6)$$

Using a procedure described in (Fisenko & Lemberg 2014a) for the representation Planck function $B_{\tilde{\nu}}(T)$ in the frequency domain, the expression for the spectral energy density has the form

$$I_{\mathrm{G}}(\nu_1,\nu_2,T)=\sum_{i=1}^{3}G_i(l,b)\tau_i\int_{\nu_1}^{\nu_2}\varepsilon(\nu)B'_{\nu}(T_i)\mathrm{d}\nu\,. \qquad (7)$$

Here $\varepsilon(\nu)=\left(\frac{\nu}{0.9\,\mathrm{THz}}\right)^2$. The Planck function in the frequency domain is defined as follows

$$B'_{\nu}(T)=\frac{8\pi h}{c^3}\frac{\nu^3}{e^{\frac{h\nu}{k_{\mathrm{B}}T}}-1}\,. \qquad (8)$$

The total emissivity in the finite range of frequencies $\nu_1\le\nu\le\nu_2$ has the following structure:

$$\varepsilon(T)=\sum_{i=1}^{3}\left(\frac{I_G(\nu_1,\nu_2,T_i)}{\int_{\nu_1}^{\nu_2}B'_{\nu}(T_i)\mathrm{d}\nu}\right)\;, \qquad (9)$$

where $\int_{\nu_1}^{\nu_2}B'_{\nu}(T)\mathrm{d}\nu$ is the total energy density of the blackbody radiation. According to (Fisenko & Lemberg 2014 a), the latter has the following structure:

$$\int_{\nu_1}^{\nu_2}B'_{\nu}(T_i)\mathrm{d}\nu=\frac{48\pi k_{\mathrm{B}}{}^4}{c^3h^3}\sum_{i=1}^{3}T_i{}^4[P_3(x_{1i})-P_3(x_{2i})], \qquad (10)$$

where $x_{1i}=\frac{h\nu_1}{k_{\mathrm{B}}T_i}$ and $x_{2i}=\frac{h\nu_2}{k_{\mathrm{B}}T_i}$. $P_3(x)=\sum_{s=0}^{3}\frac{(x)^s}{s!}\mathrm{Li}_{4-s}(e^{-x})$ is defined using the polylogarithm function $\mathrm{Li}_{4-s}(e^{-x})=\sum_{k=1}^{\infty}\frac{e^{-kx}}{k^{4-s}}$ (Abramowitz & Stegun 1972).

According to (Landau & Lifshitz 1980), the free energy density can be represented in the form

$$f(\nu_1,\nu_2,T)=\frac{8\pi k_B}{c^3}\sum_{i=1}^{3}G_i(l,b)\,T_i\int_{\nu_1}^{\nu_2}\nu^2\varepsilon(\nu)\ln\left((1-e^{-\frac{h\nu}{k_B T_i}}\right)d\nu\ . \tag{11}$$

where $\tau_1$ and $T_1$ are the optical depth and temperature for the warm component, $\tau_2$ and $T_2$ for the intermediate-temperature component, and $\tau_3$ and $T_3$ for the very cold component.

Using Eq. 11 the thermodynamic functions of the galactic far infrared radiation are defined as follows

1. The entropy density

$$s=-\frac{\partial f}{\partial T} \tag{12}$$

2. Pressure

$$P=-f \tag{13}$$

3. Heat capacity at constant volume per unit volume

$$c_V=\left(\frac{\partial I_G(\nu_2,\nu_2,T)}{\partial T}\right)_V \tag{14}$$

4. The number density of photons $n=\dfrac{N}{V}$

$$n=\frac{8\pi}{c^3\nu_0^2}\sum_{i=1}^{3}G_i(l,b)\,\tau_i\int_{\nu_1}^{\nu_2}\frac{\nu^4}{e^{\frac{h\nu}{k_B T_i}}-1}d\nu \tag{15}$$

## 3. Radiative properties of the galactic far infrared radiation

Let us now construct an expression for the total energy density of the galactic far infrared radiation. In this case, according to Eq. 7, we need to calculate the integral of the following form:

$$I_0(\nu_1,\nu_2,T)=G(l,b)\,\tau\int_{\nu_1}^{\nu_2}\varepsilon(\nu)B_\nu'(T\ )\mathrm{d}\nu \tag{16}$$

Using the analytical expression for the spectral emissivity Eq. 4, after integration, we obtain

$$I_0(\nu_1,\nu_2,T)=G(l,b)\,a\,\tau\,T^6\left[P_5(x_1)-P_5(x_2)\right]. \tag{17}$$

Then, Eq. 7 for the total energy density of the galactic far infrared radiation takes the form

$$I_{\mathrm{G}}(\nu_1,\nu_2,T) = a\sum_{i=1}^{3} G_i(l,b)\tau_i T_i^{6}\left[P_5(x_{1i}) - P_5(x_{2i})\right]. \tag{18}$$

Here $a = \dfrac{960\pi k_{\mathrm{B}}^{6}}{c^3 h^5 \nu_0^2} = 7.4890\times10^{-18}\ \mathrm{J\,m^{-3}K^{-6}}$.

In the semi-infinite range $0 \le \nu \le \infty$, since $P_5(0) = \mathrm{Li}_6(1) = \xi(6) = \dfrac{\pi^6}{945}$ and $P_5(\infty) = 0$, the Eq. 18 simplifies to

$$I_{\mathrm{G}}(\nu_1,\nu_2,T) = a'\sum_{i=1}^{3} G_i(l,b)\ \tau_i\ T_i^{6}\ , \tag{19}$$

where $a' = \dfrac{64\pi^7 k_{\mathrm{B}}^{6}}{63 c^3 h^5 \nu_0^2} = 7.6189\times10^{-18}\,\mathrm{J\,m^{-3}K^{-6}}$.

The total radiation power per unit area is defined as

$$I'_{\mathrm{G}}(\nu_1,\nu_2,T) = \frac{c}{4} I_G(\nu_1,\nu_2,T)\,, \tag{20}$$

and in accordance with Eqs. 18 and 19, has the following structure:

1. Finite range of frequency $\nu_1 \le \nu \le \nu_2$

$$I'_{\mathrm{G}}(\nu_1,\nu_2,T) = b\sum_{i=1}^{3} G_i(l,b)\tau_i T_i^{6}\left[P_5(x_{1i}) - P_5(x_{2i})\right]\ , \tag{21}$$

where $b = \dfrac{240\pi k_{\mathrm{B}}^{6}}{c^2 h^5 \nu_0^2} = 5.6129\times10^{-10}\ \mathrm{W\,m^{-2}K^{-6}}$.

2. Semi-infinite range $0 \le \nu \le \infty$

$$I'_{\mathrm{G}}(0,\infty,T) = b'\sum_{i=1}^{3} G_i(l,b)\ \tau_i\ T_i^{6}\ , \tag{22}$$

where $b' = \dfrac{16\pi^7 k_{\mathrm{B}}^{6}}{63 c^2 h^5 \nu_0^2} = 5.7102\times10^{-10}\,\mathrm{W\,m^{-2}K^{-6}}$.

As seen, in accordance with Eq. 21 and Eq. 22, the temperature dependence of the total radiation power per unit area differs from well-known Stefan-Boltzmann law $I(T) \propto T^4$.

Using Eq. 10 and Eq. 18 for the total emissivity Eq. 9, we obtain

$$\varepsilon(T) = A\sum_{i=1}^{3} \frac{G_i(l,b)\ \tau_i\ T_i^{2}\left[P_5(x_{1i}) - P_5(x_{2i})\right]}{[P_3(x_{1i}) - P_3(x_{2i})]}\ , \tag{23}$$

where $A = \frac{20k_B^2}{h^2 v_0^2} = 0.01071\text{K}^{-2}$.

In the semi-infinite frequency range $0 \le \nu \le \infty$ , Eq. 23 is simplified to

$$\varepsilon(T) = A' \sum_{i=1}^{3} G_i(l,b)\tau_i\, T_i^2 \tag{24}$$

with $A' = \frac{40\pi^2 k_B^2}{21h^2 v_0^2} = 0.01007\,\text{K}^{-2}$.

In Table 1 and Table 2, the values for the radiative functions for the warm, intermediate-temperature and very cold components as well as a sum of components are presented. As seen in Table 1 and Table 2, the finite frequency range from 0.15 to 2.88 THz covers a substantial portion of the total galactic continuous spectrum.

It is well-known that for the construction of cosmological models of radiative transfer in the inner Galaxy the generalized Stefan-Boltzmann law is used in the following form:

$$I^{\text{S-B}}(T) = \varepsilon(T)\sigma T^4 . \tag{25}$$

Here $\sigma = 5.6704\times10^{-8}$ J m$^{-2}$s$^{-1}$K$^{-4}$is the Stefan-Boltzmann constant.

However, the temperature dependence of $\varepsilon(T)$ should be known. The three-component model allows to obtain analytical expressions for the total emissivity $\varepsilon(T)$. In accordance with Eq. 24, and assuming $G_i = 1$, we obtain:

a) Warm dust component

$$\varepsilon(T) = c\,T^2 , \tag{26}$$

where $c = 1.7522\times10^{-7}$ K$^{-2}$.

b) Intermediate-temperature component

$$\varepsilon(T) = d\,T^2 , \tag{27}$$

where $d = 1.007\times10^{-6}$ K$^{-2}$.

c) Very cold component

$$\varepsilon(T) = e\,T^2 , \tag{28}$$

where $e = 1.2386\times10^{-6}$ K$^{-2}$.

Then, generalized Stefan-Boltzmann laws, which can be used for modelling the radiative transfer in the warm dust region, intermediate-temperature component dust region, and very cold dust region in our Galaxy, have the following structure:

a) Warm dust component

$$I^{\text{S-B}}(T) = \sigma' T^6 , \tag{29}$$

where $\sigma' = 9.8831 \times 10^{-15}\ \text{J m}^{-2}\ \text{s}^{-1}\text{K}^{-6}$. $\sigma'$ can be called as the generalized Stefan-Boltzmann constant for the warm component.

b) Intermediate-temperature dust component

$$I^{\text{S-B}}(T) = \sigma'' T^6 , \tag{30}$$

where $\sigma'' = 5.6799 \times 10^{-14}\ \text{J m}^{-2}\ \text{s}^{-1}\text{K}^{-6}$. $\sigma''$ can be called as the generalized Stefan-Boltzmann constant for the intermediate-temperature component.

c) Very cold dust component

$$I^{\text{S-B}}(T) = \sigma''' T^6 , \tag{31}$$

where $\sigma''' = 6.99862 \times 10^{-14}\ \text{J m}^{-2}\ \text{s}^{-1}\text{K}^{-6}$. $\sigma'''$ can be called as the generalized Stefan-Boltzmann constant for the very cold component.

In conclusion, let us apply these results to calculate the radiative parameters of the galactic far infrared radiation (photon gas) for our Galaxy. The total radiation power emitted from the surface $S'$ of the Galaxy has the form

$$I_{\text{G total}}(T) = S' I'_G(v_1, v_2, T) . \tag{32}$$

The continuous spectrum measured by FIRAS *COBE* instrument allows us to calculate the total radiation power (total luminosity) emitted from a surface of our Galaxy. Indeed, the calculated total radiation power $I'_G(v_1, v_2, T)$ emitted from unit area is presented in Table 1 and Table 2. According to Eq. 25, we need to know the value of the surface area $S'$ of our Galaxy. It is well known that the Milky Way Galaxy is a stellar disc with a diameter *d* of about 30 kpc and about 0.7 kpc wide *h* (Martinez & Saar 2001). The disc has a spiral structure and populated by interstellar dust, interstellar gas and metal-rich stars. The surface area of the Galaxy can be determined as

$$S' = 2\pi \left(\frac{d}{2}\right)^2 + \pi h d\ . \tag{33}$$

Now let's assume that all thermal emitted objects (stars etc.), as well as interstellar dust, warm dust, intermediate-temperature dust, and very cold dust in Milky Way Galaxy are located on the surface of the Galaxy, and they uniformly distributed. The stars emit the thermal radiation isotropically in all direction. Then, Eq. 32 takes the form

1. Finite range $0.15\,\mathrm{THz} \le \nu \le 2.88\,\mathrm{THz}$

$$I_{\mathrm{G\ total}} = 2\pi\left(\frac{d}{2}\right)h\left[1+\frac{d}{2h}\right]\ I'_G(\nu_1,\nu_2,T) \tag{34}$$

2. Semi-infinite range $0 \le \nu \le \infty$

$$I_{\mathrm{G\ total}} = 2\pi\left(\frac{d}{2}\right)h\left[1+\frac{d}{2h}\right]\ I'_G(0,\infty,T) \tag{35}$$

Using the following relationship $1\,\mathrm{kpc} \cong 3.086\times 10^{19}\ \mathrm{m}$ the value for a surface area of the Galaxy Eq. 33 is equal to

$$S' = 1.4085\times 10^{42}\ \mathrm{m}^2\ . \tag{36}$$

Then, for the total radiation power (total luminosity) emitted from surface area $S'$ , we have

1. Finite range $0.15\,\mathrm{THz} \le \nu \le 2.88\,\mathrm{THz}$

$$I'_{\mathrm{G\ total}}(T) = 9.1530\times 10^{35}\ \mathrm{W} \tag{37}$$

2. Semi-infinite range $0 \le \nu \le \infty$

$$I'_{\mathrm{G\ total}}(T) = 1.0482\times 10^{36}\ \mathrm{W} \tag{38}$$

These values are different from the value of $I'_{\mathrm{G\ total}}(T) \approx 1.0\times 10^{37}\ \mathrm{W}$ given in the literature (http://www.daviddarling.info/encyclopedia/G/Galaxy.html). This difference can be explained by using the spectral emissivity $\varepsilon(\nu)$ in our calculations.

The volume of the disk of the Milky Way Galaxy can be calculated using the following expression

$$V_{\mathrm{Galaxy}} = \pi\, h\left(\frac{d}{2}\right)^2 = 1.4534\times 10^{61}\,\mathrm{m}^3 \tag{39}$$

Then, according to Eq. 18 and 19 and Table 1 and Table 2, the total energy density of photons produced by our galaxy has the following values:

1. Finite range $0.15\,\mathrm{THz} \le v \le 2.88\,\mathrm{THz}$

$$I_{\mathrm{G\,total}}(T) = V_{\mathrm{Galaxy}}\, I_{\mathrm{G}}(T) = 1.2062 \times 10^{47}\ \mathrm{J} \tag{40}$$

2. Semi-infinite range $0 \le v \le \infty$

$$I_{\mathrm{G\,total}}(T) = V_{\mathrm{Galaxy}}\, I_{\mathrm{G}}(T) = 1.4431 \times 10^{47}\ \mathrm{J} \tag{41}$$

## 4. Thermodynamics of the galactic far infrared radiation

To construct the thermodynamics of the far infrared galactic radiation let's start with the calculation of the free energy. According to (Landau & Lifshitz 1980) and Eq.11, the free energy density is given by formula

$$f(v_1, v_2, T) = \frac{8\pi k_{\mathrm{B}}}{c^3 v_0^2} \sum_{i=1}^{3} G_i(l,b)\tau_i T_i \int_{v_1}^{v_2} v^4 \ln\left((1 - e^{-\frac{hv}{k_{\mathrm{B}} T_i}}\right) dv. \tag{42}$$

Introducing the variable $x = \dfrac{hv}{k_{\mathrm{B}} T}$, and integrating by parts, we obtain

$$f(v_1, v_2, T) = -B \sum_{i=1}^{3} G_i(l,b)\, \tau_i\, T_i^6 \left[ (P_5(x_{1i}) - P_5(x_{2i})) - \frac{1}{120}(x_{1i}^5\ \mathrm{Li}_1(e^{-x_{1i}}) - x_{2i}^5\ \mathrm{Li}_1(e^{-x_{2i}})) \right], \tag{43}$$

where $B = \dfrac{192\pi k_{\mathrm{B}}^6}{c^3 h^5 v_0^2} = 1.4978 \times 10^{-18}\ \mathrm{J\ m^{-3}\ K^{-6}}$.

In $0 \le v \le \infty$ frequency range Eq. 43 simplifies to

$$f(v_1, v_2, T) = -B' \sum_{i=1}^{3} G_i(l,b)\, \tau_i\, T_i^6\ . \tag{44}$$

Here $B' = \dfrac{64\pi^7 k_{\mathrm{B}}^6}{315 c^3 h^5 v_0^2} = 1.5238 \times 10^{-18}\ \mathrm{J\ m^{-3}\ K^{-6}}$.

The entropy density $s = -\dfrac{\partial f}{\partial T}$ in the finite frequency range has the following structure:

$$s = C \sum_{i=1}^{3} G_i(l,b)\, \tau_i\, T_i^5 \left[ (P_5(x_{1i}) - P_5(x_{2i})) - \frac{1}{720}(x_{1i}^5\ \mathrm{Li}_1(e^{-x_{1i}}) - x_{2i}^5\ \mathrm{Li}_1(e^{-x_{2i}})) \right], \tag{45}$$

where $C = \dfrac{1152\pi k_B^6}{c^3 h^5 v_0^2} = 8.9868\times10^{-18}\,\mathrm{J\,m^{-3}\,K^{-5}}$ .

In semi-infinite frequency range Eq. 45 transforms to

$$s = C'\sum_{i=1}^{3} G_i(l,b)\,\tau_i\,T_i^5 \quad . \tag{46}$$

Here $C' = \dfrac{128\pi^7 k_B^6}{105 c^3 h^5 v_0^2} = 9.1427\times10^{-18}\,\mathrm{J\,m^{-3}\,K^{-5}}$ .

The pressure is defined by formula

$$P = -f = D\sum_{i=1}^{3} G_i(l,b)\,\tau_i\,T_i^6\left[(P_5(x_{1i}) - P_5(x_{2i})) - \frac{1}{120}(x_{1i}^5\,\mathrm{Li}_1(e^{-x_{1i}}) - x_{2i}^5\,\mathrm{Li}_1(e^{-x_{2i}}))\right], \tag{47}$$

where $D = \dfrac{192\pi k_B^6}{c^3 h^5 v_0^2} = 1.4978\times10^{-18}\,\mathrm{J\,m^{-3}\,K^{-6}}$.

In the semi-infinite range Eq. 47 simplifies to

$$P = D'\sum_{i=1}^{3} G_i(l,b)\,\tau_i\,T_i^6 \,. \tag{48}$$

Here $D' = \dfrac{64\pi^7 k_B^6}{315 c^3 h^5 v_0^2} = 1.5238\times10^{-18}\,\mathrm{J\,m^{-3}\,K^{-6}}$.

Knowledge of the total energy density Eq. 18 allows us to construct the heat capacity at constant volume per unit volume $c_V = \left(\dfrac{\partial I_0(v_2, v_2, T)}{\partial T}\right)_V$ in the form

$$c_V = E\sum_{i=1}^{3} G_i(l,b)\,\tau_i\,T_i^5\left[(P_5(x_{1i}) - P_5(x_{2i})) + \frac{1}{720}(x_{1i}^6\,\mathrm{Li}_0(e^{-x_{1i}}) - x_{2i}^6\,\mathrm{Li}_{01}(e^{-x_{2i}}))\right], \tag{49}$$

where $E = \dfrac{5760\pi k_B^6}{c^3 h^5 v_0^2} = 4.4934\times10^{-17}\,\mathrm{J\,m^{-3}\,K^{-5}}$.

In semi-infinite frequency range Eq. 49 is presented as follows

$$c_V = E'\sum_{i=1}^{3} G_i(l,b)\,\tau_i\,T_i^5 \,. \tag{50}$$

Here $E' = \dfrac{128\pi^7 k_B^6}{21 c^3 h^5 v_0^2} = 4.5713\times10^{-18}\,\mathrm{J\,m^{-3}\,K^{-5}}$.

Using the formula for the number density of photons

$$n = \frac{8\pi}{c^3 \nu_0^2} \sum_{i=1}^{3} G_i(l,b)\, \tau_{\mathrm{i}} \int_{\nu_1}^{\nu_2} \frac{\nu^4}{e^{\frac{h\nu}{k_{\mathrm{B}} T_i}} - 1} d\nu \ , \tag{51}$$

after integration, we obtain

$$n = F \sum_{i=1}^{3} G_i(l,b)\, \tau_i\, T_i^{5} \left[ P_4(x_{1i}) - P_4(x_{2i}) \right], \tag{52}$$

where $F = \frac{192\pi k_{\mathrm{B}}^5}{c^3 h^5 \nu_0^2} = 1.0848 \times 10^5 \ \mathrm{m^{-3}\ K^{-5}}$ .

In the $0 \le \nu \le \infty$ range Eq. 52 takes the form

$$n = F' \sum_{i=1}^{3} G_i(l,b)\, \tau_i\, T_i^{5} \ . \tag{53}$$

Here $F' = \frac{64\pi^6 k_{\mathrm{B}}^5}{315 c^3 h^5 \nu_0^2} = 1.1037 \times 10^5 \ \mathrm{m^{-3}\ K^{-5}}$.

By definition (Landau & Lifshitz 1980), the chemical potential density $\mu = \left( \frac{\partial f}{\partial n} \right)_{T,V}$ , as clearly seen from Eq. 35, is equal to zero.

These expressions are used in computer calculations for the three-component galactic model. In Table 1 and Table 2, the calculated values for the radiative and thermodynamic functions of the warm intermediate-temperature, and cold components, and also a sum of three-components are represented in the finite and semi-infinite frequency ranges. To calculate the values for the radiative and thermodynamic functions we used the following data: a) warm component - the mean temperature $T_1 = 17.72\ \mathrm{K}$ and optical depth $\tau_1 = 1.74 \times 10^{-5}$; b) the intermediate-temperature component - the mean temperature $T = 14\,\mathrm{K}$ and mean optical depth $\tau_2 = 1.00 \times 10^{-4}$ c) very cold component - the mean temperature $T_1 = 6.75\ \mathrm{K}$ and optical depth $\tau_2 = 1.23 \times 10^{-4}$.

In conclusion, let us calculate the thermodynamic properties of the galactic far infrared radiation (photon gas) for our Galaxy.

Using Eq. 39 and the values for the number density of photons presented in Table 1 and Table 2, for the thermodynamic properties of photon gas in the far-infrared frequency range $0.15\,\mathrm{THz} \le \nu \le 2.88\,\mathrm{THz}$ and the semi-infinite range, we obtain

1. The total number of photons $N_{\mathrm{G\ total}}$ in our Galaxy

   a) Finite range $0.15\,\mathrm{THz} \le v \le 2.88\,\mathrm{THz}$

$$N_{\mathrm{G\ total}} = V_{Galaxy} n = 1.3229 \times 10^{68} \tag{54}$$

   b) Semi-infinite range $0 \le v \le \infty$

$$N_{\mathrm{G\ total}} = V_{Galaxy} n = 1.3780 \times 10^{68} \ . \tag{55}$$

2. The total entropy $S_{\mathrm{G\ total}}$ of photon gas

   a) Finite range $0.15\,\mathrm{THz} \le v \le 2.88\,\mathrm{THz}$

$$S_{\mathrm{G\ total}} = s V_{Galaxy} = 1.0181 \times 10^{46}\,\mathrm{J\,K^{-1}} \tag{56}$$

   b) Semi-infinite range $0 \le v \le \infty$

$$S_{\mathrm{G\ total}} = s V_{Galaxy} = 1.1415 \times 10^{46}\,\mathrm{J\,K^{-1}} \ . \tag{57}$$

3. The total free energy $F_{\mathrm{G\ total}}$ of photon gas

   c) Finite range $0.15\,\mathrm{THz} \le v \le 2.88\,\mathrm{THz}$

$$F_{\mathrm{G\ total}} = f\, V_{Galaxy} = -2.7019 \times 10^{46}\,\mathrm{J} \tag{58}$$

   d) Semi-infinite range $0 \le v \le \infty$

$$F_{\mathrm{G\ total}} = f\, V_{Galaxy} = -2.8862 \times 10^{46}\,\mathrm{J} \ . \tag{59}$$

4. The total heat capacity at constant value $C_{V\,\mathrm{G\ total}}$

   e) Finite range $0.15\,\mathrm{THz} \le v \le 2.88\,\mathrm{THz}$

$$C_{V\,\mathrm{G\ total}} = c_V\, V_{Galaxy} = 4.5464 \times 10^{46}\,\mathrm{J\,K^{-1}} \tag{60}$$

   f) Semi-infinite range $0 \le v \le \infty$

$$C_{V\,\mathrm{G\ total}} = c_V\, V_{Galaxy} = 5.7076 \times 10^{46}\,\mathrm{J\,K^{-1}} \ . \tag{61}$$

5. Total pressure $P_{\mathrm{G\ total}}$

g) Finite range $0.15\,\mathrm{THz} \le \nu \le 2.88\,\mathrm{THz}$

$$P_{\mathrm{G\ total}} = pV_{Galaxy} = 2.7019\times10^{46}\,\mathrm{J} \quad (62)$$

a) Semi-infinite range $0 \le \nu \le \infty$

$$P_{\mathrm{G\ total}} = pV_{Galaxy} = 2.8862\times10^{46}\,\mathrm{J}\ . \quad (63)$$

In Table 3, the radiative and thermodynamic properties of the galactic far infrared radiation (photon gas) in the finite and semi-infinite frequency ranges for the Milky Way Galaxy are presented.

In conclusion, it is important to note the following. The obtained results can be used to describe the radiative and thermodynamic properties for any external galaxies in the universe. In favor of what has been said it is worth noting the work of Spinoglio (2005). In this work, a two-component model with the emissivity law in the form $\varepsilon \sim \nu^2$ to fit the observed thermal continuum spectrum for the Seyfert 2 galaxy NGC 1068 is proposed. As a result, the expressions obtained in this paper are applicable for calculating the radiative and thermodynamic properties of this Galaxy. This topic will be point of discussion in subsequent publication.

## 5. Astrophysical parameters

Now let as calculate astrophysical parameter for galactic far infrared photons. In accordance with the table of astrophysical constants and parameters (Groom 2013), three fundamental constants and parameters for the monopole spectrum, such as the present day CMB temperature, entropy density/Boltzmann constant, and number density of CMB photons are presented. The parameter of the entropy density/Boltzmann constant, for example, has the form

$$\frac{s}{k_{\mathrm{B}}} = 2.8912\left(\frac{T}{T_0}\right)^3\ \mathrm{cm}^{-3}, \quad (64)$$

where $T_0 = 2.72548$ K is present day CMB temperature.

Let us assume that the optical depth varies with temperature. Then, as in the case of a monopole spectrum, the expressions for the entropy density/Boltzmann constant and the number density of galactic far infrared photons can be considered as additional parameters to the astrophysical ones presented in table (Groom 2013). In accordance with Eq. 46 and Eq. 53, the analytical expressions for the parameter of entropy density/Boltzmann constant and number density of galactic far infrared photons are presented in the form

1. Entropy density/Boltzmann constant

a) Warm dust component

$$\frac{s}{k_{\mathrm{B}}} = 20.1304\left(\frac{\tau}{\tau_{01}}\right)\left(\frac{T}{T_{01}}\right)^{5}\ \mathrm{cm}^{-3} \tag{65}$$

b) Intermediate-temperature component

$$\frac{s}{k_{\mathrm{B}}} = 35.6151\left(\frac{\tau}{\tau_{02}}\right)\left(\frac{T}{T_{02}}\right)^{5}\ \mathrm{cm}^{-3} \tag{66}$$

c) Very cold component

$$\frac{s}{k_{\mathrm{B}}} = 1.1413\left(\frac{\tau}{\tau_{03}}\right)\left(\frac{T}{T_{03}}\right)^{5}\ \mathrm{cm}^{-3}\quad . \tag{67}$$

2. Number density of galactic far-infrared photons

a) Warm dust component

$$n = 3.3551\left(\frac{\tau}{\tau_{01}}\right)\left(\frac{T}{T_{01}}\right)^{5}\ \mathrm{cm}^{-3} \tag{68}$$

b) Intermediate-temperature component

$$n = 5.9358\left(\frac{\tau}{\tau_{02}}\right)\left(\frac{T}{T_{02}}\right)^{5}\ \mathrm{cm}^{-3} \tag{69}$$

c) Very cold component

$$n = 0.1902\left(\frac{\tau}{\tau_{03}}\right)\left(\frac{T}{T_{03}}\right)^{5}\ \mathrm{cm}^{-3}. \tag{70}$$

Here $\tau_{0i}$ and $T_{0i}$ are the present day optical depths and temperatures. Their values are: a) $\tau_{01} = 1.74 \times 10^{-5}$ and $T_{01} = 17.72$ K for the warm component; b) $\tau_{02} = 1.00 \times 10^{-4}$ and $T_{02} = 14$ K for the intermediate-temperature component; c) $\tau_{03} = 1.23 \times 10^{-4}$ and $T_{03} = 6.75$ K for the very cold component.

## 6. Conclusions

In this paper, using the three-component model to describe the thermal continuum spectra of galactic far infrared radiation, the analytical expressions for the temperature dependences of the radiative and thermodynamic functions, such as the total radiation power per unit area, total energy density, total emissivity, number density of photons, Helmholtz free energy density, entropy density, heat capacity at constant volume and pressure are obtained.

The radiative and thermodynamic properties of our Galaxy in the finite frequency range from 0.15 THz to 2.88 THz and semi-infinite range are calculated. In this case, the following parameters were used for the warm, intermediate-temperature, and cold components: a) warm component - the mean temperature $T_1 = 17.72\,\mathrm{K}$ and mean optical depth $\tau_1 = 1.77 \times 10^{-5}$; b) intermediate-temperature component – the mean temperature $T = 14\,\mathrm{K}$ and mean optical depth $\tau_2 = 1.00 \times 10^{-4}$; and c) very cold component - the mean temperature $T_1 = 6.75\,\mathrm{K}$ and mean optical depth $\tau_1 = 1.23 \times 10^{-4}$. The calculated values are presented in Table 1 and Table 2.

Knowing the dependence of the temperature *T* and the frequency *v* on redshift z (Sunyaev and Zel'dovich, 1980), allows us to study the thermodynamic and radiative state of our galaxy, many years ago with the help of analytical expressions derived in this paper.

Under the assumption that all the thermal emitted objects (stars etc.), as well as interstellar dust, the warm dust, intermediate-temperature dust, and very cold dust uniformly distributed on the surface of our Galaxy, the total radiation power received from a surface of the Galaxy in the finite frequency range from 0.15 THz to 2.88 THz and semi-infinite range are calculated. Their values are $I_{\mathrm{G\,total}}(T) = 3.3757 \times 10^{35}\,\mathrm{W}$ and $I_{\mathrm{G\,total}}(T) = 4.2286 \times 10^{35}\,\mathrm{W}$, accordingly. The calculated values for other radiative and radiative properties for the Milky Way Galaxy are presented in Table 3.

The total numbers of photons in finite and semi-infinite ranges in our Galaxy are obtained. For the finite frequency range $0.15\,\mathrm{THz} \le v \le 2.88\,\mathrm{THz}$, we have $N_{\mathrm{G\,total}} = 1.3229 \times 10^{68}$. For the semi-infinite range of frequencies the total number of photons is $N_{\mathrm{G\,total}} = 1.3780 \times 10^{68}$ .

The thermodynamic properties of galactic far-infrared photons, such as the total Helmholtz free energy, total entropy, total heat capacity at constant volume and total pressure for the Galaxy are calculated. The value for the total entropy in the finite frequency range $0.15\,\mathrm{THz} \le \nu \le 2.88\,\mathrm{THz}$ is $S_{\mathrm{G\ total}} = 1.0181 \times 10^{46}\,\mathrm{J\,K^{-1}}$. For the semi-finite frequency range**,** we have $S_{\mathrm{G\ total}} = 1.1415 \times 10^{46}\,\mathrm{J\,K^{-1}}$ . As for the total pressure the values are: $P_{\mathrm{G\ total}} = 2.7019 \times 10^{46}\,\mathrm{J}$ - finite frequency range; $P_{\mathrm{G\ tota}} = 2.8862 \times 10^{46}\,\mathrm{J}$ - semi-infinite range. Other thermodynamic values for our Galaxy are presented in Table 3.

The generalized Stephan-Boltzmann law for the warm, intermediate-temperature, and cold components are constructed and presented by Eqs. 29-31. Their temperature dependence has the structure $I^{\mathrm{S\text{-}B}}(T) = \sigma' T^6$, which differs from well-known the Stephan-Boltzmann law ( $I^{\mathrm{S\text{-}B}}(T) = \sigma T^4$ ). These results are important for the construction of cosmological models of radiative transfer in our Galaxy.

The results of the present paper allow estimating the contribution of the radiative and thermodynamic properties of the photon gas to the total radiation of other particles (protons, alpha and beta particles etc.).

The analytical expressions obtained in this paper can be useful for calculating the radiative and thermodynamic properties of any of the Galaxy, for which a two-component model is a good approximation. The Seyfert 2 galaxy NGC 1068 is one of examples where the two-component model can be applied.

This topic will be point of discussion in subsequent publication.

**Acknowledgments**

The authors sincerely thank Dr. A. Zhuk for discussions.

| **Quantity** | **Warm Component** $\nu_1 \le \nu \le \nu_2$ | **Intermediate-Temperature Component** $\nu_1 \le \nu \le \nu_2$ | **Very Cold Component** $\nu_1 \le \nu \le \nu_2$ | **Sum of Components** $\nu_1 \le \nu \le \nu_2$ |
|---|---|---|---|---|
| $I_G(\nu_1, \nu_2, T)$ [J m$^{-3}$] | $3.2536\times10^{-15}$ | $5.3285\times10^{-15}$ | $8.8509\times10^{-17}$ | $8.6706\times10^{-15}$ |
| $I'_G(\nu_1, \nu_2, T)$ [W m$^{-2}$] | $2.4385\times10^{-7}$ | $3.9936\times10^{-7}$ | $6.6336\times10^{-9}$ | $6.4984\times10^{-7}$ |
| $\varepsilon\ (T)$ | $4.5812\times10^{-5}$ | $1.8636\times10^{-4}$ | $5.8750\times10^{-5}$ | $2.9092\times10^{-4}$ |
| $f$ [J m$^{-3}$] | $-7.3035\times10^{-16}$ | $-1.1110\times10^{-15}$ | $-1.7618\times10^{-17}$ | $-1.8590\times10^{-15}$ |
| $s$ [J m$^{-3}$ K$^{-1}$] | $2.2483\times10^{-16}$ | $4.5996\times10^{-16}$ | $1.5723\times10^{-17}$ | $7.0051\times10^{-16}$ |
| $P$ [J m$^{-3}$] | $7.3035\times10^{-16}$ | $1.1110\times10^{-15}$ | $1.7618\times10^{-17}$ | $1.8590\times10^{-15}$ |
| $c_V$ [J m$^{-3}$ K$^{-1}$] | $9.2633\times10^{-16}$ | $2.1230\times10^{-15}$ | $7.8748\times10^{-17}$ | $3.1281\times10^{-15}$ |
| $n$ [m$^{-3}$] | $3.0494\times10^{6}$ | $5.8605\times10^{6}$ | $1.9229\times10^{5}$ | $9.1022\times10^{6}$ |

**Table 1** Calculated values of the radiative and thermodynamic state functions for the galactic far-infrared spectrum in 0.15 – 2.88 THz frequency interval. $G(l,b) = 1$. a) $T_1 = 17.72$ K and $\tau_1 = 1.74 \times 10^{-5}$ for the warm component; b) $T_2 = 14$ K and $\tau_2 = 1.00 \times 10^{-4}$ for the intermediate-temperature component; c) $T_3 = 6.75$ K and $\tau_3 = 1.23 \times 10^{-4}$ for the very cold component.

| Quantity | Warm Component<br><br>$0 \le \nu \le \infty$ | Intermediate-Temperature Component<br>$0 \le \nu \le \infty$ | Very Cold Component<br><br>$0 \le \nu \le \infty$ | Sum of Components<br><br>$0 \le \nu \le \infty$ |
|---|---|---|---|---|
| $I_G(0,\infty,T)$<br>[J m$^{-3}$] | $4.1041\times10^{-15}$ | $5.7367\times10^{-15}$ | $8.8638\times10^{-17}$ | $9.9294\times10^{-15}$ |
| $I'_G(0,\infty,T)$<br>[W m$^{-2}$] | $3.0760\times10^{-7}$ | $4.2995\times10^{-7}$ | $6.6433\times10^{-9}$ | $7.4419\times10^{-7}$ |
| $\varepsilon\ (T)$ | $5.5020\times10^{-5}$ | $1.9738\times10^{-4}$ | $5.6436\times10^{-5}$ | $3.0884\times10^{-4}$ |
| $f$<br>[J m$^{-3}$] | $-8.2083\times10^{-16}$ | $-1.1473\times10^{-15}$ | $-1.7728\times10^{-17}$ | $-1.9858\times10^{-15}$ |
| $s$<br>[J m$^{-3}$ K$^{-1}$] | $2.7793\times10^{-16}$ | $4.9172\times10^{-16}$ | $1.5758\times10^{-17}$ | $7.8541\times10^{-16}$ |
| $P$<br>[J m$^{-3}$] | $8.2083\times10^{-16}$ | $1.1473\times10^{-15}$ | $1.7728\times10^{-17}$ | $1.9858\times10^{-15}$ |
| $c_V$<br>[J m$^{-3}$ K$^{-1}$] | $1.3897\times10^{-15}$ | $2.4586\times10^{-15}$ | $7.8789\times10^{-17}$ | $3.9271\times10^{-15}$ |
| $n$ [m$^{-3}$] | $3.3551\times10^{6}$ | $5.9358\times10^{6}$ | $1.9022\times10^{5}$ | $9.4811\times10^{6}$ |

**Table 2** Calculated values of the radiative and thermodynamic state functions for the galactic far-infrared spectrum in the semi-infinite frequency interval. $G_i(l,b)=1$. a) $T_1 = 17.72$ K and $\tau_1 = 1.74\times10^{-5}$ for the warm component; b) $T_2 = 14$ K and $\tau_2 = 1.00\times10^{-4}$ for the intermediate-temperature component; c) $T_3 = 6.75$ K and $\tau_3 = 1.23\times10^{-4}$ for the very cold component.

| **Quantity** | **Three-Component Model** $0.15\,\mathrm{THz} \le \nu \le 2.88\,\mathrm{THz}$ | **Three-Component Model** $0 \le \nu \le \infty$ |
|---|---|---|
| $I_{\mathrm{G\ total}}$ [J] | $1.2062\times10^{47}$ | $1.4431\times10^{47}$ |
| $I'_{\mathrm{G\ total}}$ [W] | $9.1530\times10^{35}$ | $1.0482\times10^{36}$ - |
| $F_{\mathrm{G\ total}}$ [J] | $-2.7019\times10^{46}$ | $-2.8862\times10^{46}$ |
| $S_{\mathrm{G\ total}}$ [J K$^{-1}$] | $1.0181\times10^{46}$ | $1.1415\times10^{46}$ |
| $P_{\mathrm{G\ total}}$ [J] | $2.7019\times10^{46}$ | $2.8862\times10^{46}$ |
| $C_{V\,\mathrm{G\ total}}$ [J K$^{-1}$] | $4.5464\times10^{46}$ | $5.7076\times10^{46}$ |
| $N_{\mathrm{G\ total}}$ | $1.3229\times10^{68}$ | $1.3780\times10^{68}$ |

**Table 3** Radiative and thermodynamic properties of the photon gas for our Galaxy. $G_i(l,b) = 1$.